# A Qualitative Study of IT Students' Skill Development: Comparing Online and Face-to-Face Learning Environments

HUGO SILVA

Faculty of Sciences and Technology, University of Coimbra, Portugal, hugosilva@dei.uc.pt

Each student has specific characteristics and learning preferences, that reflect on each type of learning environment, online or face-to-face. Understanding these differences is crucial for educators to create learning environments that can inspire and engage students. This qualitative study explores and tries to better understand, specifically the IT student's experiences and perceived skills development in online and face-to-face learning environments, while trying to address the question: "Regarding online and face-to-face learning environments, in IT, how do students experience and assess their skill development in one learning environment compared to the other?". Using a social constructive paradigm, the purpose of the research is to focus as much as possible on the student's views of the situation and how their perspectives and experiences shape the perception of developed skills. Data was collected through semi-structured interviews by focusing on the student and asking for their personal experience on skill development through online and face-to-face learning environments. The data analysis strategy adopts the grounded theory approach, using a systematic procedure. The results suggest that face-to-face learning may develop a better communication and collaborative skills more effectively while experiencing a synchronous interaction, where online learning may strength in self-regulation and adaptability skills because of the independence and flexibility it provides. This study produces two grounded theories that explain how different IT learning environments influence the development of student's specific skills, that can contribute to pedagogical discussions on optimizing hybrid learning experiences.

CCS CONCEPTS • Qualitative study • Grounded theory • IT learning environments

**Additional Keywords and Phrases:** Online learning, Face-to-face learning, IT student experience and assessment, Skill development

## 1 INTRODUCTION

In recent years, there has been a significant shift in educational delivery modes, especially with the increase of online learning platforms. The COVID-19 pandemic accelerated this transition, highlighting the need for effective online teaching strategies. This shift is particularly relevant for Information Technology (IT) students, whose field demands a balance of theoretical and practical knowledge, with a focus on hands-on skills. However, there remains a gap in understanding how these skill sets are developed in online versus face-to-face (F2F) learning environments. Understanding these dynamics is critical to ensuring IT graduates are well-prepared for the workforce, regardless of the learning modality.

This study seeks to explore how IT students perceive and develop key skills – such as problem-solving, teamwork, and technical proficiency – in online and F2F learning environments. Specifically, we aim to compare the strengths, challenges, and overall effectiveness of each modality in fostering skill development. To address these objectives, we conducted a qualitative study involving semi-structured interviews with IT students who had experienced both online and

F2F learning. Data was collected and analyzed using the grounded theory approach, using a systematic procedure to uncover common patterns, perceptions, and insights into their learning experiences. The findings of this study have significant implications for educators, institutions, and policymakers. With the increasing integration of online learning, it is vital to ensure that this mode of education supports the same level of skill acquisition as traditional F2F methods. By understanding student's experiences and the factors that facilitate or hinder skill development, we can design better curricula, adopt effective teaching strategies, and ensure equitable learning outcomes for IT students across all delivery formats. This, in turn, contributes to a more competent and adaptable IT workforce.

The present paper proposes a qualitative study on IT student's experiences and perceived skills development in online and F2F learning environments. The main contributions of this paper are:

- Proposal and comparison of two theories, one for each learning environment, while performing a grounded theory qualitative study.

The remainder of this paper is organized as follows. Section 2 presents related work on qualitative study of IT Students' Skill Development, while comparing online and F2F learning environments. Section 3 presents the methodology and Section 4 the data collection. Section 5 describes the data analysis, while finally Section 6 discusses the main findings.

## 2  MINIMAL STATE-OF-THE-ART

This section describes a minimal literature review of related works, namely qualitative studies, when possible, of IT student's experiences and perceived skills development in online and F2F learning environments.

Mather, et al. (2018) [1] analyzed 313 student perspectives on online and F2F learning in a single course, comparing performance, challenges, satisfaction, and achievements. Online learning was valued for its flexibility, particularly by mature students with external commitments or those far from campus. However, online learners reported dissatisfaction with delayed faculty feedback and engagement difficulties. In contrast, F2F learners appreciated the immediate interaction with faculty and peers, which enhanced understanding and provided prompt feedback. Both groups faced challenges with group work, with online learners experiencing more difficulty in virtual collaboration. The study suggests strategies such as clear guidelines, role assignments, and support systems to improve group project outcomes, emphasizing the importance of tailored approaches to leverage the strengths of each learning mode.

Lewohl (2023) [2] examined a new teaching strategy to improve student engagement in a second-year neurobiology course. The strategy integrated lectures, interactive workshops with technology, and online resources. Students who attended in-person classes performed better and reported significant benefits, though those using class recordings for scheduling flexibility achieved comparable results. This suggests that offering multiple access modes, including recordings and digital resources, supports performance and enhances engagement and satisfaction. While the study focused on one course, the findings may have broader applications across various disciplines.

Pinto-Llorente, et al. (2018) [3] explored students' perceptions of essential traits for success in a blended learning environment, focusing on 91 learners over three months. The study found that strong English proficiency and digital skills are crucial for studying an English degree in a blended format. Students highlighted the importance of mastering synchronous and asynchronous tools for communication and assessments. Key personal attributes identified for success included maturity, self-reflection, organization, perseverance, innovation, and collaboration within the e-learning community.

Almanar, et al. (2020) [4] conducted a qualitative study using online questionnaires to examine students' responses to distance learning during the COVID-19 pandemic. The study explored learning experiences, media, and tools used by educators. Positive impacts included improvements in computational skills, autonomy, and critical thinking. However, negative impacts were attributed to issues with connectivity, lack of compatible devices, limited basic computational skills, and challenges in student-teacher communication.

Wright (2017) [5] examined student perceptions of an online lesson in an English grammar course compared to F2F lessons, focusing on motivation and interest. Most students found F2F lessons more motivating and engaging due to better understanding facilitated by direct interaction with lecturers and peers. In contrast, students who preferred online lessons highlighted the advantages of speed, convenience, and flexibility in time and location.

## 3 METHODOLOGY

This qualitative study explores and tries to better understand, specifically the IT student's experiences and perceived skills development in online and F2F learning environments, while trying to address the research question: "Regarding online and F2F learning environments, in IT, how do students experience and assess their skill development in one learning environment compared to the other?". Using a social constructive paradigm, the purpose of the research is to focus as much as possible on the student's views of the situation and how their perspectives and experiences shape the perception of developed skills. The data analysis strategy will adopt the grounded theory approach approach from Cresswell (2013) [6], using a systematic procedure, while using open coding to identify initial categories, axial coding to explore relationships between categories surrounding the core phenomena in a visual model, and selective coding to develop hypotheses from the model to assemble a story describing the interrelationship of categories in the model.

The research questions will focus on: 'What was the process?' and 'How did it unfold?'. After, more detailed questions will be made for the axial coding phase: 'What was central to the process?' (the core phenomenon); 'What influenced or caused this phehomenon to occur?' (causal conditions); 'What strategies were employed during the process?' (strategies); 'What effect occurred?' (consequences). This will allow to gather rich and detailed information to generate a theory for each learning environment, illustrated in a figure, to explain how IT students experience the skill development on both environments.

## 4 DATA COLLECTION

Data was collected from eight IT students across bachelor, master, and PhD programs at three universities: the Department of Informatics Engineering (DEI) of the Faculty of Science and Technology at the University of Coimbra, the Department of Informatics and Systems Engineering (DEIS) of the Coimbra Institute of Engineering at the Polytechnic Institute of Coimbra, and the Department of Arts and Technologies (DAT) of the Superior School of Education of Coimbra at the Polytechnic Institute of Coimbra. Most participants (75%) were from DEIS, where courses were held after working hours, while 12.5% each were from DEI (during working hours) and DAT (after working hours). All participants have experienced both online and F2F learning formats, particularly during and after the COVID-19 pandemic in 2020.

The data was collected through semi-structured interviews using a 24-question open-ended questionnaire divided into four sections, designed to explore participants' experiences and perceptions of skill development in online and F2F IT learning environments:

1. **Main Phenomena**: Focused on understanding participants' experiences, challenges, and motivations in IT learning environments;
2. **Causal Conditions**: Investigated factors influencing their experiences, including external characteristics and lecturer methodologies;

3. **Strategies**: Explored personal strategies used to adapt and overcome challenges, such as time management, task prioritization, and communication approaches;
4. **Consequences**: Examined which skills participants felt were most developed in each learning environment.

This structure aimed to provide in-depth insights into the factors shaping skill development in IT education. Data collection was completed over two weeks using a structured questionnaire sent via email to participants. This approach was chosen for its convenience and to reduce potential intimidation or conditioning, allowing participants to reflect and provide more thoughtful, in-depth responses to the open-ended questions. The study generated a comprehensive dataset of 81 pages of textual data, offering detailed insights into how students perceived and navigated their online and F2F learning environments. It highlighted the factors influencing their skill development in each modality. The data was analyzed using grounded theory, as detailed in the subsequent analysis section.

## 5 DATA ANALYSIS

As mentioned previously in the methodology section, the data strategy will adopt the grounded theory approach from Cresswell (2013) [6], using initially open coding to identify initial categories, followed by axial coding to explore relationships between categories surrounding the core phenomena in a visual model, and finally the selective coding to develop hypotheses from the model to assemble a story describing the interrelationship of categories in the model. The final purpose is to, for each learning environment, do a comparison of the two grounded theories by pointing the strengths and weaknesses of each learning environment.

### 5.1 Face-to-face learning environment

In the F2F learning environment, initial open coding identified key categories such as "Direct interaction with classmates and professors", "Access to physical resources" (e.g., libraries and offices), and "Concentration", with students emphasizing the benefits of group dynamics and being in a dedicated physical workspace, as stated: "In F2F environment is the group dynamics and the possibility of discussing more immediately with my colleagues and supervisors and the feeling of being in a physical space dedicated to work also creates an environment of greater concentration". Participants highlighted that F2F settings facilitated the development of skills, originating categories like "Communication", "Collaboration", "Teamwork", "Idea exchange", "Brainstorming", and "Problem-solving", as stated: "In the F2F environment, the development of non-technical skills, such as communication, collaboration and teamwork, was central. Direct interaction with colleagues and advisors provides a more dynamic space for discussing ideas, brainstorming and solving problems together. Additionally, participating in F2F meetings and team projects helped me improve my teamwork and communication skills".

Further categories, such as "Empathy," "Commitment with team members," and "Sharing knowledge," emerged, reflecting the interpersonal and professional skills gained. For instance, one participant noted that these skills are vital for future teamwork, idea presentation, and interactions with clients and colleagues, as stated: "These skills help me a lot in the future because communication is essential for working in a team, presenting ideas and sharing knowledge and dealing with clients, colleagues and leaders".

During the process, recurrent patterns were detected in F2F learning. For example, regarding the motivations, social interaction and collaboration were the primary sources of motivation for students. Being physically present created a sense of shared goals and community, as stated: "What motivated me most was the social interaction and the fact that we were all working towards the same goals", "In the F2F environment, my motivation comes mainly from interacting with others", and "In a F2F environment, group/class dynamics are important to increase willingness to participate". Also, patterns regarding human connection and direct engagement that contributed to students feeling more involved in the learning process as stated: "The F2F environment allows direct and immediate interaction with my colleagues, which facilitates the exchange of ideas", and "Direct interaction with the professor and colleagues creates a more engaging environment". Time management was a recurring challenge, particularly for group work, which required coordinating schedules and balancing commitments: "In person, it wasn't always easy to manage time, especially when it involved group work where we had to juggle everyone's schedules". Presentation anxiety was another issue faced by students in F2F contexts: "A challenging moment can be the nervousness of presenting a project or idea in front of the class". Non-technical skills in F2F learning were perceived as more effective for developing interpersonal skills such as communication, teamwork, collaboration, and leadership, as stated: "Communication, leadership, and group collaboration skills were better developed in the F2F environment", and "In the F2F environment we can better develop our communication and non-technical skills".

Through axial coding, interrelationships among these categories were identified, forming a grounded theory model (shown in Figure 1) that captures the core dynamics of skill development in the F2F learning environment.

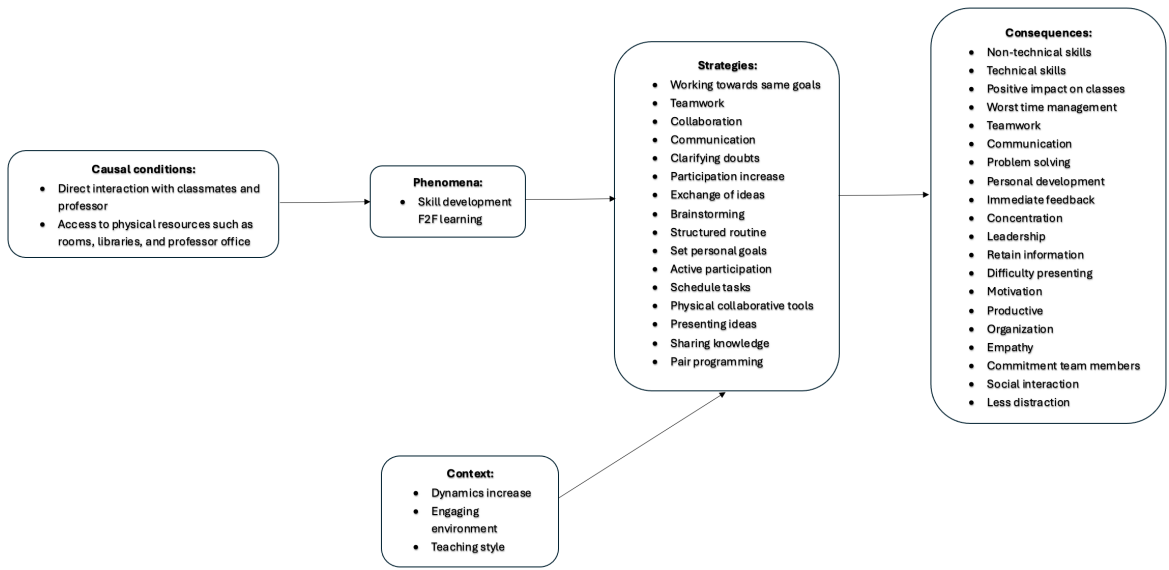

Figure 1 – Theoretical Model for IT Students' Skill Development in F2F Learning Environment.

The analysis identified five main groups of categories: Causal Conditions, Phenomena, Context, Strategies, and Consequences, which together describe the process of skill development in the F2F learning environment. Two key Causal Conditions were identified: "Direct interaction with classmates and professors" and "Access to physical resources" (e.g., classrooms, libraries, and professor offices). These factors motivated students and created an environment conducive to concentration and collaboration. For instance, participants emphasized the value of interpersonal interaction and the availability of physical resources in enhancing their learning experience, as stated: "In a F2F context, I am motivated by the fact that I can interact with my classmates", and "In the F2F environment, the factors that most influenced my experience were the direct interaction with professors and colleagues and the physical environment, that is, the location of classes and access to university resources, such as rooms and libraries that created an environment of concentration". The Strategies employed were influenced by the Causal Conditions, the resulting Phenomena (skill development in F2F learning), and the Context (e.g., increased dynamics, engaging environment, teaching style). Strategies included teamwork, collaboration, communication, increased participation, exchange of ideas, setting personal goals, scheduling tasks, and pair programming, all of which facilitated skill development. The strategies had both positive and negative Consequences. Positive consequences included the improvement of non-technical skills, better communication, enhanced concentration, motivation, leadership development, and a positive impact on classes. On the downside, some participants reported poor time management and difficulty in presenting ideas as challenges. This framework demonstrates how causal factors and strategies influence skill development in F2F learning, resulting in a mix of personal and academic growth along with some challenges.

The selective coding process generated the following hypotheses: skill development in F2F learning is driven by physical presence, direct interaction with classmates and professors, and access to physical resources. Strategies such as teamwork, collaboration, communication, idea exchange, brainstorming, and setting personal goals were influenced by the increased dynamics and engaging environment of F2F learning. These strategies resulted in several positive outcomes, including improved non-technical skills, better communication, teamwork spirit, personal development, increased concentration, and a positive impact on classes. However, there were some downsides, such as poor time management and difficulties in presenting ideas. Overall, the physical presence in F2F learning enhances social interaction but may reduce time management efficiency due to logistical challenges like commuting.

## 5.2 Online learning environment

In the online learning environment, initial open coding identified key categories such as "Discipline," "Time management," "Organization," "Flexibility," "Technical skills," and "Problem solving". Participants emphasized the importance of autonomy and personal management, highlighting the flexibility to learn at their own pace as a driver for developing organizational and discipline skills. Access to online resources like tutorials and platforms such as Stack Overflow and GitHub facilitated independent problem-solving and technical skill development, as stated: "In the online environment, autonomy and personal management have become even more crucial. The flexibility to learn at my own pace helped me develop greater organizational and discipline skills. Technical skills are also more easily developed in this environment, due to access to a wide range of online resources, such as tutorials, and support materials that I can consult at any time. Problem solving has become more focused on learning independently, sometimes using platforms like Stack Overflow or GitHub".

Additional categories included "Proactivism," "Digital literacy", "Responsibility", and "Prioritization". Students noted that online learning helped them become more proactive in finding solutions, adapt to changes, and develop communication and technology-related skills. Managing deadlines and prioritizing tasks further strengthened their sense of responsibility and time management abilities. Overall, the online environment fostered independence and adaptability while encouraging efficient use of digital tools and resources, as stated: "I believe that I developed autonomy and management of my personal time, I became more flexible to new changes and to be more proactive in finding solutions. I think that online helped to develop some communication skills and made us understand and develop skills in using technologies that we were not used to. Solving problems more proactively. Time management is also important to meet deadlines and gain responsibility. In both environments, I learned to prioritize tasks and manage my time better".

During the process, recurrent patterns were detected in online learning. For example, regarding the motivations, flexibility and autonomy were consistently highlighted as motivating factors. Students valued the ability to manage their time better and learn at their own pace. For example, as stated: "I am motivated by the fact that I can manage my time better", "The flexibility and availability of materials make the environment more accessible", and "In the online environment, my motivation is more internal because I have more autonomy over my time". Regarding challenges, distraction and monotony were common challenges. The lack of physical presence often created a sense of isolation and reduced engagement, as stated: "In an online environment, it is easier to get distracted, especially on theoretical topics", "Online classes tend to be much more monotonous, since most students have their cameras turned off", and "The sudden lack of interaction made me feel a certain isolation and difficulty concentrating". Also, technical skills, such as programming and problem solving, were often perceived as being developed in online learning, due to access to a wide range of online resources and the ability to practice independently, as stated: "Technical skills, such as programming and problem solving, were more developed in the online environment", and "The freedom to study at your own pace and the use of digital tools help in the development of technical skill".

Recurrent patterns were also found regarding positive and negative aspects. For the positive aspects, flexibility and accessibility of content were praised for enabling students to balance academic and personal responsibilities, as stated: "Online, time management and self-discipline are key", and "The flexibility to learn at my own pace helped me develop greater organizational and discipline skills". On the negative aspects, the solitary nature of online learning often reduced interpersonal connection and collaboration, making it feel less dynamic, as stated: "The online regime is solitary, more in the style of save yourself if you can", "Being remote makes work more lonely and less dynamic", and "This experience is difficult to fully replicate online, even with video or chat tools, as physical presence has a different energy".

The next phase, axial coding, focused on exploring the interrelationships between categories around the core phenomena, resulting in the grounded theory model shown in Figure 2. The analysis revealed six main groups: Causal Conditions, Phenomena, Context, Intervening Conditions, Strategies, and Consequences. Three key Causal Conditions for the Phenomena "Skill development in online learning" emerged: "Access to online platforms and content", "Flexibility", and "Pandemic." Participants highlighted the impact of accessible materials, flexible schedules, and the challenges posed by fully online classes during the pandemic, as stated "In the online environment, the factors that most influence my experience are the materials available and the flexible schedule. Flexibility and autonomy and ease of access to content, especially recorded classes. For the online environment, a challenging moment was during the pandemic, when classes became completely online".

The Strategies used for skill development were shaped by these causal conditions, the resulting Phenomena (skill development), the Context (e.g., asynchronous tasks, use of technology, solitary atmosphere, and monotony), and Intervening Conditions like internet connectivity. These factors influenced how students adapted and developed skills in the online learning environment. The Strategies employed by participants included categories like discipline, setting personal goals, using collaborative technological tools, requesting feedback, scheduling frequent meetings, learning independently, and improving time management.

These strategies led to both positive and negative Consequences. On the positive side, participants experienced improvements in technical skills (e.g., programming), time management, autonomy, problem-solving, responsibility, digital literacy, and prioritization. However, negative outcomes included distraction, a negative impact on classes, poor interpersonal interaction, decreased concentration, procrastination, and fatigue.

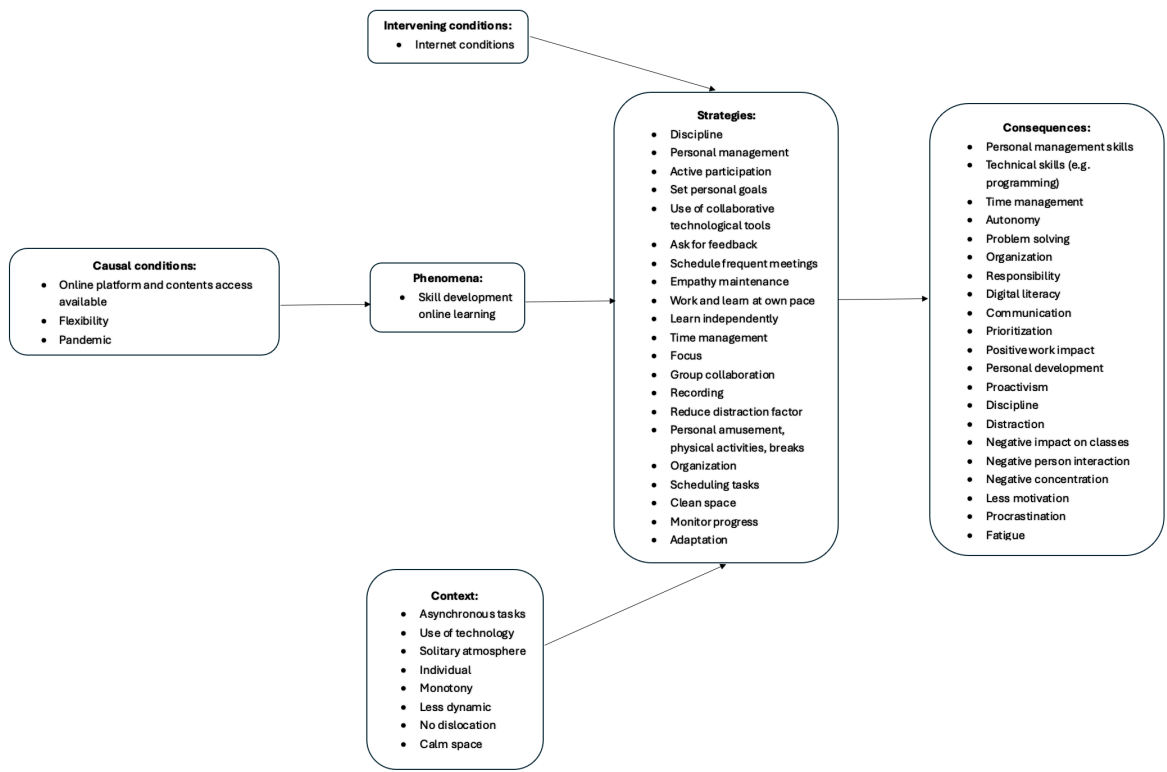

Figure 2 – Theoretical Model for IT Students' Skill Development in Online Learning Environment.

The selective coding process led to the following hypotheses: skill development in online learning is influenced by the availability of online platforms and content, flexibility, and the pandemic. Strategies such as discipline, active participation, setting personal goals, and using collaborative technological tools were employed, influenced by factors like internet conditions, asynchronous tasks, technology use, monotony, and the lack of physical presence. These strategies resulted in improvements in technical skills, time management, autonomy, problem-solving, and responsibility. However, negative consequences included poor interpersonal interactions, decreased concentration, lower motivation, and procrastination. Overall, while online learning enhanced technical skills, time management and autonomy, it also limited personal interaction and concentration due to the more isolated nature of the environment.

### 5.3 Comparing F2F and online IT skills' development grounded theories

Both F2F and online learning environments have distinct dynamics that influence skill development. The analysis of these environments reveals commonalities and differences in how students develop skills, employ strategies, and experience various consequences.

Causal Conditions:
- F2F Learning: The key causal factors were direct interaction with classmates and professors and access to physical resources such as classrooms, libraries, and professor offices. These factors fostered a motivating environment, creating a setting conducive to concentration and collaboration;

- Online Learning: The primary causal factors identified were access to online platforms and content, flexibility, and the pandemic. The flexibility of learning at one's own pace and access to materials were crucial for skill development. However, the pandemic posed challenges as it forced classes to be entirely online, impacting the overall experience.

Phenomena (Skill Development):
- F2F Learning: The skill development was greatly influenced by social interaction, collaboration, and the ability to access physical resources.
- Online Learning: The skill development was more influeced by the access to the online platform and contents.

Strategies:
- F2F Learning: The strategies employed included teamwork, collaboration, communication, and participation. These strategies were influenced by the dynamics of in-person interactions, which led to an engaging environment where students could share ideas, brainstorm, and problem-solve together. Setting personal goals and scheduling tasks were also key strategies that enhanced skill development;
- Online Learning: Strategies were influenced by asynchronous tasks, the use of technology, and the lack of direct interaction. Students employed strategies like discipline, setting personal goals, using collaborative technological tools, and learning independently.

Consequences:
- F2F Learning: Positive outcomes included improved non-technical skills, better communication, and enhanced concentration. Additionally, motivation and leadership development were commonly reported. On the downside, some students experienced poor time management and difficulties presenting ideas, likely due to external factors like commuting and class schedules;
- Online Learning: Positive outcomes included improvements in technical skills (e.g., programming), time management, autonomy, and problem-solving. The flexibility allowed students to manage their own learning pace, which improved digital literacy and prioritization. However, negative outcomes included distractions, poor interpersonal interactions, decreased concentration, procrastination, and fatigue. The lack of direct interaction and the solitary nature of online learning contributed to a sense of isolation, which impacted students' motivation and engagement.

Hypotheses and Key Insights:
- F2F Learning: Skill development was largely driven by physical presence, direct interaction, and access to resources. The engaging environment, facilitated by in-person connections, contributed to the development of more the non-technical skills, but also technical. However, logistical issues, such as commuting, could hinder time management;
- Online Learning: Skill development, mostly technical skills like programming, was influenced by the flexibility of the format and the availability of digital resources, along with the challenges posed by the pandemic. The absence of F2F interaction meant that online learning fostered more independent learning, but at the cost of personal interaction and concentration.

Through the data analysis, we observed a clear and consistent pattern in the responses related to skill perception and development across both online and F2F learning environments. This consistency in the themes and examples suggests that the data reached saturation—the point at which additional data no longer introduced new insights or themes. This reinforces the reliability and robustness of the findings, allowing me to trust the results and the conclusions drawn.

# 6 DISCUSSION

The research aimed to explore how IT students experience and assess their skill development in both online and F2F learning environments. By analyzing the factors influencing skill development and the strategies used in each environment, we gain insight into how students perceive the effectiveness of each learning context in developing technical and non-technical skills.

In line with the research question, students in the F2F learning environment reported more favorable experiences when it came to the development of non-technical skills such as communication, teamwork, and leadership. The physical presence of peers and instructors, as well as direct interaction, was a key factor in this skill development. The ability to engage in group discussions, brainstorming, and problem-solving in real-time contributed to the dynamic and engaging nature of F2F classes. This aligns with the research question's inquiry into how F2F interactions contribute to skill development: students found that the direct interaction with classmates and professors created an environment conducive to collaboration and the exchange of ideas, which helped enhance their interpersonal and non-technical skills. However, while F2F environments promoted stronger social interaction and group collaboration, students did face challenges, notably in time management, due to external factors such as commuting and the fixed schedule of classes. This reflects a potential drawback when comparing F2F learning with online environments, which offer more flexibility.

When examining the online learning environment, the findings revealed that flexibility and access to online resources were the driving forces behind technical skill development and autonomy. Students in online courses were able to learn at their own pace, which fostered a sense of independence and improved time management skills. The ability to access materials at any time, such as recorded lectures and tutorials, allowed students to develop technical skills, particularly in programming and other IT-related tasks. This directly answers the research question regarding how students assess their skill development in online learning: students perceived that the self-paced nature and flexibility of online learning environments helped them develop technical competencies and problem-solving abilities. However, the lack of direct interaction in online learning environments negatively impacted the development of non-technical skills like communication and teamwork. Furthermore, students expressed challenges related to distractions, procrastination, and fatigue, which negatively impacted their overall experience in online learning. The monotony and solitary nature of online courses created a less engaging atmosphere compared to the more dynamic F2F setting.

These insights align with the research question, demonstrating that the comparison between online and F2F learning environments provides valuable understanding into how students assess and experience skill development in the context of IT education. The blending of both environments in a hybrid learning model may offer the best of both worlds, allowing students to develop both technical and non-technical skills while enjoying the flexibility of online learning combined with the social benefits of F2F interaction.

This study adds new insights to the existing literature on IT students' experiences and skill development in online and F2F learning environments, by providing two grouded theories, and comparison between them. Key contributions include the examination on how IT students perceive skill development in both online and F2F settings, highlighting the development of technical (e.g., programming) and non-technical skills (e.g., communication, teamwork). It deepens the understanding of how direct interaction with classmates and professors in F2F environments enhances non-technical skills, such as collaboration and problem-solving. The study reinforces the benefits of flexibility in online learning, showing how it enhances autonomy, time management, and technical skills development, adding new insights into the relationship between flexibility and technical skill acquisition in IT. While earlier research acknowledged challenges in online learning, this study provides a more detailed analysis of distractions, poor interpersonal interactions, and decreased concentration in online settings, specifically for IT students. It further points out the pandemic's role in online learning, highlighting how the shift to fully online education during the pandemic increased challenges and changed skill development. The study offers a detailed look at strategies for success in both environments, with a focus on time management, goal setting, and collaboration tools in online learning, and teamwork and communication in F2F settings. Finally, it suggests the potential benefits of a hybrid learning model, integrating the strengths of both online and F2F approaches to maximize skill development.